\documentclass[twocolumn,showpacs,aps,prl]{revtex4-1}
\usepackage{verbatim} 
\usepackage{graphicx}
\usepackage{dcolumn}
\usepackage{bm}
\usepackage{array}
\usepackage{booktabs}
\usepackage{amsmath}
\usepackage{color}
\usepackage{times}
\usepackage{epstopdf}

\usepackage[T1]{fontenc}

\begin{document}
\title {\boldmath
$\mathcal R(3780)$ Resonance Interpreted as the $1^3D_1$-Wave 
Dominant State of Charmonium from Precise Measurements of the Cross Section of
$e^+e^-\rightarrow$~Hadrons
}
\author{
\begin{small}
\begin{center}
M.~Ablikim$^{1}$, M.~N.~Achasov$^{13,b}$, P.~Adlarson$^{75}$, X.~C.~Ai$^{81}$, R.~Aliberti$^{36}$, A.~Amoroso$^{74A,74C}$, M.~R.~An$^{40}$, 
Q.~An$^{71,58}$, Y.~Bai$^{57}$, O.~Bakina$^{37}$
% ...
\\
\vspace{0.2cm}
(BESIII Collaboration)\\
\vspace{0.2cm} {\it
$^{1}$ Institute of High Energy Physics, Beijing 100049, People's Republic of China\\
% ...
$^{13}$ G.I. Budker Institute of Nuclear Physics SB RAS (BINP), Novosibirsk 630090, Russia\\
% ...
$^{36}$ Johannes Gutenberg University of Mainz, Johann-Joachim-Becher-Weg 45, D-55099 Mainz, Germany\\
$^{37}$ Joint Institute for Nuclear Research, 141980 Dubna, Moscow region, Russia\\
% ...
$^{40}$ Liaoning Normal University, Dalian 116029, People's Republic of China\\
% ...
$^{57}$ Southeast University, Nanjing 211100, People's Republic of China\\
$^{58}$ State Key Laboratory of Particle Detection and Electronics, Beijing 100049, Hefei 230026, People's Republic of China\\
% ...
$^{71}$ University of Science and Technology of China, Hefei 230026, People's Republic of China\\
% ...
$^{74}$ University of Turin and INFN, (A)University of Turin, I-10125, Turin, Italy; (B)University of Eastern Piedmont, I-15121, Alessandria, Italy; (C)INFN, I-10125, Turin, Italy\\
$^{75}$ Uppsala University, Box 516, SE-75120 Uppsala, Sweden\\
% ...
$^{81}$ Zhengzhou University, Zhengzhou 450001, People's Republic of China\\
\vspace{0.2cm}
% ...
$^{b}$ Also at the Novosibirsk State University, Novosibirsk, 630090, Russia\\
% ...
}
\end{center}
  \vspace{0.4cm}
  \end{small}
}
\noaffiliation
\vspace{0.0cm}

\begin{abstract}
We report the precise measurements of the cross section of $e^+e^-\rightarrow$~hadrons 
at center-of-mass energies from 3.645 to 3.871 GeV.
We thereby perform the most precise study of the cross sections 
and find a complex system composed of three resonances of 
$\mathcal R(3760)$, $\mathcal R(3780)$, and $\mathcal R(3810)$.
For the first time, we measure the $\mathcal R(3810)$
electronic width to be $(19.4 \pm 7.4 \pm 12.1)$~eV. 
For the $\mathcal R(3760)$ resonance, we measure the mass to be 
$(3751.9\pm 3.8\pm 2.8)$ ~MeV/$c^2$, the total width to be 
$(32.8 \pm 5.8 \pm 8.7)$~MeV, and the electronic width to be 
$(184\pm 75\pm 86)$~eV.
For the $\mathcal R(3780)$ resonance, we measure its mass to be 
$(3778.7\pm 0.5\pm 0.3)$~MeV/$c^2$, total width to be 
$(20.3 \pm 0.8 \pm 1.7)$~MeV, and electronic width to be 
$(265\pm 67\pm 83)$~eV.
Forty-seven years ago, the $\psi(3770)$ resonance was discovered, and 
was subsequently interpreted as the $1^3D_1$-wave dominant state 
of charmonium. However, our analysis of the total-hadron cross sections 
indicates that the $\psi(3770)$ is not a single state, 
but a complex system composed of the $\mathcal R(3760)$, $\mathcal R(3780)$, 
and $\mathcal R(3810)$ resonances. 
Among these, we interpret the $\mathcal R(3780)$ is a
resonance dominated by the $1^3D_1$ charmonium state.
\end{abstract}

\maketitle

In 1977, the MARK-I Collaboration discovered a single resonance 
denoted the $\psi(3770)$ in $e^+e^-\!\rightarrow$~hadrons
at center-of-mass (c.m.) energies from 3.73 to 3.87~GeV, 
and interpreted it as the $1^3D_1$-wave dominant state of charmonium~\cite{LGW,DLCO}.
Twenty years ago, the BES Collaboration reported the observation of seven occurrences 
of the nonopen-charm (nOC) hadron decay 
$\psi(3770)\rightarrow J/\psi{\pi^+}{\pi^-}$~\cite{hep_ex_0307028v1}.
Because of the fact that the mass of the $\psi(3770)$ resonance is higher than the open-charm (OC)
threshold, this observation overturned the conventional knowledge
that mesons containing charm quarks with masses above the OC-pair thresholds 
decay entirely to OC final states via the strong interaction. 
The observation of this
decay was comfirmed two~\cite{PLB605_63_Y2005} 
and three~\cite{PRL96_082004_Y2006} years latter.
This observation
inspired researchers to  study the
production and decay of meson resonances in the OC energy region,
in particular to study the line shape of $e^+e^-\rightarrow$~hadrons cross sections at energies 
from 3.65 to 3.88~GeV~\cite{PhysRevLetts97_121801_2006,PRL97_262001_Y2006}.
In this Letter, we denote these  meson resoances as $X_{\rm aboveOC}$~\cite{PRD102_112009_Year2020}, 
which encompass both heavy $c\bar{c}$ states, 
i.e.\ $\psi(3770)$, $\psi(4040)$, $\psi(4160)$, $\psi(4415)$,  
and potential non-$c\bar{c}$ states, such as four-quark states, OC-pair molecular states, 
hadro-charmonium states, and hybrid charmonium states, expected by quantum chromodynamics (QCD).
Discovery of these non-$c\bar{c}$ states would be an important validation of QCD predictions. 
Precise measurements of the resonance parameters of $X_{\rm aboveOC}$ states
provide valuable data for understanding the nature of the $X_{\rm aboveOC}$ states.

The charmonium model~\cite{Eithtin_chmonuim_prd1978} predicts that 
the $1^3D_1$ state should decay more than $99\%$ of the time into open-charm final states.
However, subsequent studies  showed that about 
$15\%$ of $\psi(3770)$ decays are into nonopen-charm final 
states~\cite{PhysLettsB641_145_2006,PhysRevLetts97_121801_2006,PhysRevD76_122002_2007,PhysLttesB659_74_2008}.
These observations suggested that there are probably undiscovered states~\cite{RongG_CPC_34_778_Y2010} 
with masses around 3.773~GeV/$c^2$, which decay into 
nonopen-charm final states with large branching fractions. 
In 2008, the BES-II experiment observed for the first time a double-peaked structure 
named the $\mathcal Rs(3770)$ ~\cite{bes2_prl_2structures}
in $e^+e^-\!\rightarrow${~hadrons}
at c.m. energies around 3.77~GeV,
which is composed of two Breit-Weigner (BW) structures 
with masses close to 3763 and 3781~MeV~\cite{bes2_prl_2structures}, 
respectively.
Reference~\cite{Dubynskiy_Voloshin_PRF78_116014_2008} 
interprets one of the two BW contributions in the double-peak structure
as a possible molecular open-charm threshold resonance. An alternative interpretation is 
a $p$-wave resonance of a four-quark ($c\bar{c}q\bar{q}$) 
state~\cite{A_De_Rujule_PRL38_317_Year1977}. 
In 2021, BESIII observed the $\mathcal R(3760)$ in 
$e^+e^-\!\rightarrow$~$J/\psi{X}$ with a 
significance of 5.3$\sigma$~\cite{di-strcures_JPsiX_arXiv_2012.04186v1},
hence establishing the double-peak structure as two resonances,
and ruling out the explanations that an enhancement 
of the cross section for $e^+e^-\!\rightarrow$~{hadrons} at 
c.m. energies near 3.76 GeV~\cite{bes2_prl_2structures} is due to the 
$e^+e^-\!\rightarrow$~$D\bar{D}$ continuum process (cnt$D\bar{D}$)
interfering with the decay $\psi(3770)\!\rightarrow$~$D\bar{D}$,
or due to $\psi(3686)\!\rightarrow$~$D\bar{D}$ decays at energies
above the $D\bar{D}$ threshold~\cite{KEDR}.
Recently, BESIII observed three resonant structures around 3.77 GeV,
labeled as $\mathcal R(3760)$, $\mathcal R(3780)$, and $\mathcal R(3810)$
in $e^+e^-$$\rightarrow$~{non-OC hadrons}~\cite{PhysRevLett132_191902_2024},
but could not measure their electronic widths.
All of these observations bring into question the conventional interpretation 
that the $\psi(3770)$ is the $1^3D_1$-wave dominant state of charmonium
and indicate the necessity of rediscovering this state.

The measurements of the parameters of the $\psi(3770)$ resonance
that are used in the determination of world average 
values~\cite{LGW,
DLCO,
MRK2,
PhysRevLetts97_121801_2006,
BaBar_PRD76_111105R_Year2007,
Bell,BES2_PLB660_315,
PhysLettsB652_238_2007,
KEDR}
do not consider the effects of  $\mathcal R(3760)$ and $\mathcal R(3810)$ decays.
It is possible, therefore, that the world-average value of the $\psi(3770)$ parameters 
may be biased. To overcome this problem, it is important to 
measure and analyze the total cross sections for $e^+e^-\!\rightarrow${~hadrons} 
with high accuracy. Such measurements will both improve the understanding 
of the known states, and help to 
interpret one of the 
$\mathcal R(3760)$, $\mathcal R(3780)$, and $\mathcal R(3810)$ as
the $1^3D_1$-wave 
dominant state of charmonium.

In this Letter, we report 
measurements
 of the total cross sections for
$e^+e^-$$\rightarrow$~hadrons 
at c.m.\ energies
from 3.645 to 3.871~GeV, 
measurements of the $\mathcal R(3760)$, $\mathcal R(3780)$, 
and $\mathcal R(3810)$ resonance parameters, the first measurement 
of the $\mathcal R(3810)$ electronic width, and  
interpretation of each of the three resonances as a specific
physical state.
The data samples used in the measurements were collected 
at 42 c.m.\ energies with the BESIII detector at the BEPCII accelerator 
in 2010. The total integrated luminosity of the data sample is 75.5 pb$^{-1}$.

The BESIII detector is described in detail in Ref.~\cite{bes3}.
The detector response is studied using samples of Monte Carlo (MC)
events which are simulated with a {\sc Geant4}-based~\cite{geant4}
detector simulation software package.
Simulated samples for all $q\bar q$ vector states (i.e. $u\bar u$, $d\bar d$,
$s\bar s$, and $c \bar c$ resonances)
and their decays to hadrons are generated using the MC event generators
{\sc kkmc}~\cite{kkmc},
{\sc evtgen}~\cite{BesEvtGen} and
{\sc lundcharm}~\cite{LundCharm_CJC_et_al_2000}.
Possible sources of background  are estimated with MC simulated
events generated using {\sc kkmc}~\cite{kkmc}, 
and the event generators {\sc babayaga}~\cite{babayaga} and {\sc twogam}~\cite{twogam}.
The detection efficiency is determined using six different  
MC simulations to describe the process 
$e^+e^-\rightarrow$~hadrons:
(i) $e^+e^-\!\rightarrow$~light-hadrons continuum processes
including lower-mass resonances (CPLMR)
with masses below 2~GeV/$c^2$,
(ii) $J/\psi \rightarrow$~hadrons, 
(iii) $\psi(3686) \rightarrow$~hadrons,
(iv) $\psi(3770) \rightarrow D^0\bar{D^0}$,
(v) $\psi(3770) \rightarrow D^+ D^-$, and
(vi) $\psi(3770) \rightarrow$~nOC.
These events include initial and final-state radiation (ISR/FSR).

\begin{figure}[]
\centering
\includegraphics[width=0.23\textwidth,height=0.17\textwidth]{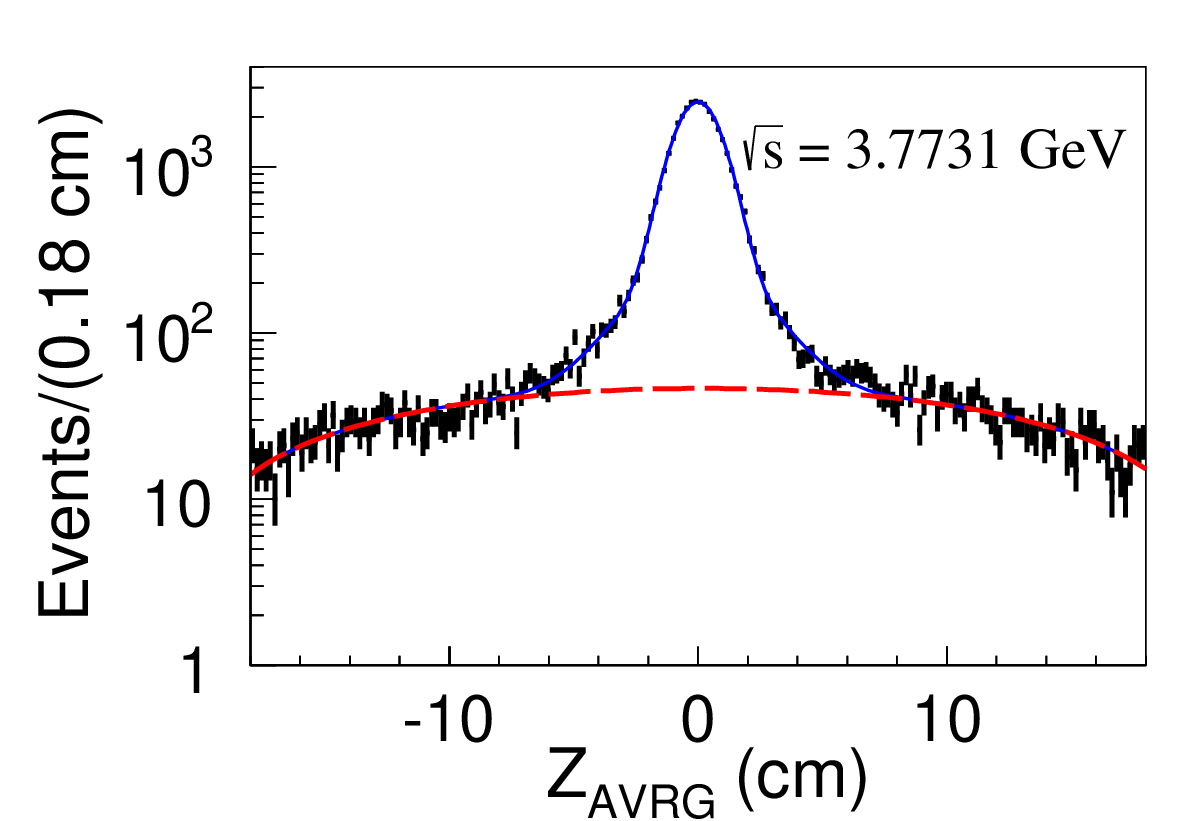}
\includegraphics[width=0.23\textwidth,height=0.17\textwidth]{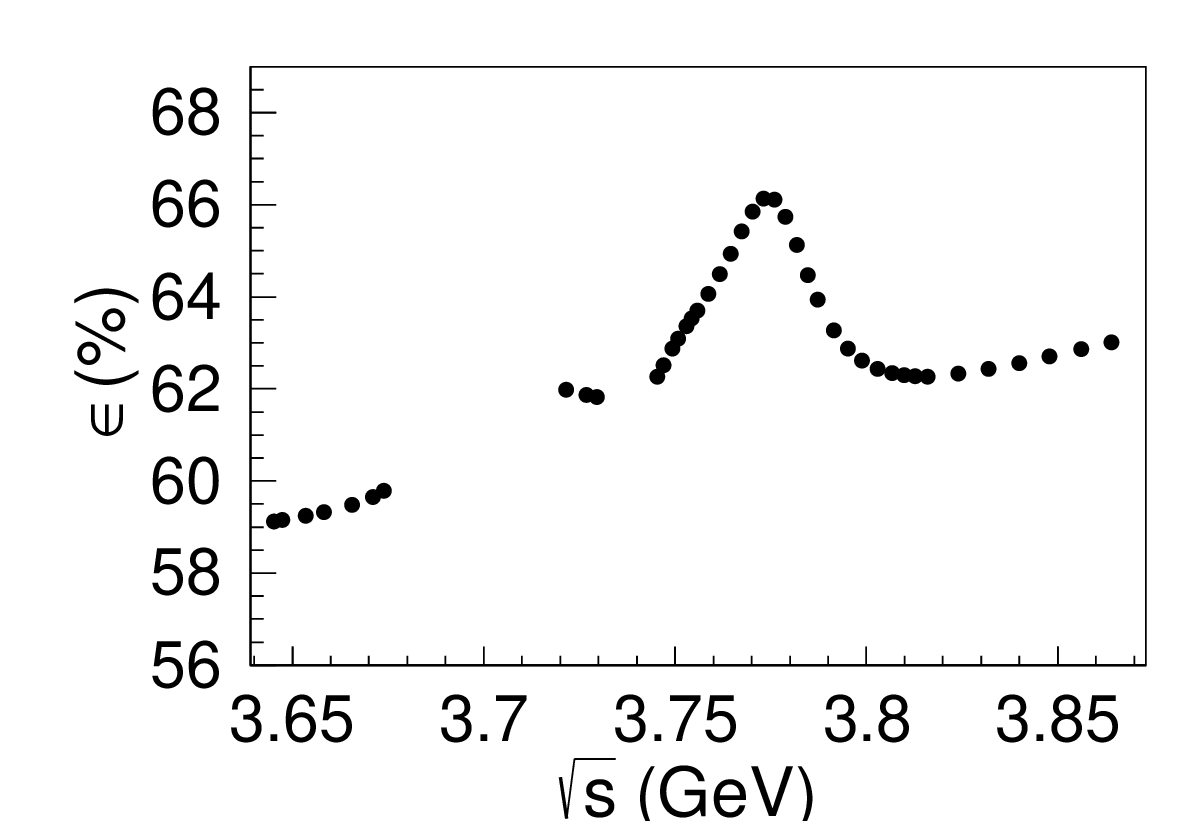}
 \caption{
Left: distribution of the averaged $z$ position 
of charged tracks for the events selected 
from the data collected at $\sqrt{s}=3.7731$ GeV,
where the dots with error bars represent data, the solid line is the best
fit and the dashed line is the fitted background shape.
Right: weighted average efficiency $\epsilon$ for selection 
of the events for $e^+e^-\rightarrow {\rm hadrons}$
at the 42 energies.
  }
 \label{fig:Emumu_over_Ecm_ecm4420MeV}
\end{figure}

Inclusive hadronic events of $e^+e^-\rightarrow$~hadrons are selected
from the charged and neutral-particle final states.
In order to 
suppress
background contributions from  
$e^+e^-\rightarrow \ell^+\ell^-$ ($\ell=e,\mu, \tau$), 
each event is required to have more than two charged tracks
in the final state~\cite{PhysLettsB641_145_2006}. 
These tracks are required to satisfy the selection criteria
described in Ref.~\cite{PhysRevLett132_191902_2024}.
To separate beam-associated background events, we determine the longitudinal position of the event
vertex, i.e.\ the average $z$ position of charged tracks in the direction of the beam line.
Figure~\ref{fig:Emumu_over_Ecm_ecm4420MeV} (left)
shows the distribution of the average $z$
for the events selected from the data sample collected 
at the c.m.\ energy $\sqrt{s}=3.7731$~GeV.
The distribution is fitted with a double-Gaussian function (using a common 
peak)
to describe the signal contribution, and a second-order 
Chebychev polynomial function to describe the background. 
The fit yields the number of candidates for hadronic events to be 
$N_{\rm had,fit}^{\rm obs}=35235\!\pm\!225$.
We determine the number
of hadronic candidate events for 
the data samples collected at the other energies in a similar manner.

The fitted signal yield $N_{\rm had,fit}^{\rm obs}$ is 
contaminated by a peaking background from several sources, e.g.
$e^+e^-\rightarrow (\gamma)\ell^+\ell^-$,
$e^+e^-\rightarrow (\gamma)e^+e^-\ell^+\ell^-$, and
$e^+e^-\rightarrow (\gamma)e^+e^-{\rm hadrons}$.
Knowing the cross sections for these processes, and using large samples of MC simulated events,
the number of peaking background events is estimated to be 
$N_{\rm b}=1547\pm 4$, where the uncertainty is 
due to the statistical uncertainty of the integrated luminosity of the data sample.
Subtracting $N_{\rm b}$ from $N^{\rm obs}_{\rm had,fit}$ yields 
$N^{\rm obs}_{\rm had}=(33688\pm 225)$
signal events.

The weighted average (WA) efficiency $\epsilon$ for the selection of hadronic events
is determined using $\epsilon=\sum^{6}_{1}w_{i} \epsilon_{i}$, where $w_{i}$
is the ratio of the number of the hadronic events selected
from the $i$th MC simulation component over the total number 
of the MC simulated hadronic events from the aforementioned 
six MC simulation components, while $\epsilon_{i}$ is the 
corresponding efficiency. Since the branching fraction 
for $\psi(3770)\!\rightarrow$~nOC includes all contributions from decays
of the full structure $\mathcal Rs(3770)$~\cite{bes2_prl_2structures}, this efficiency
includes all contributions from the decays of the BW structures contained in 
$\mathcal Rs(3770)$~\cite{bes2_prl_2structures}.
Figure~\ref{fig:Emumu_over_Ecm_ecm4420MeV} (right)
shows the WA efficiencies for the 42 energies used in this analysis.

At $\sqrt{s}=3.7731$ GeV, the WA efficiency 
is $66.13\%$. The integrated luminosity of the data 
collected at $\sqrt{s}=3.7731$ GeV is $\mathcal L=(1831.63\pm 4.49)$~nb$^{-1}$. 
The signal yield  $N^{\rm obs}_{\rm had}$ divided by both the luminosity and WA efficiency
yields the 
measured 
observed cross section $\sigma^{\rm obs}_{\rm had}=(27.813\pm 0.198)$~nb,
where the uncertainty is of statistical origin and accounts for the number 
of observed events, the MC sample size and the statistical uncertainty 
of the luminosity measurement. This result 
 is in good agreement with the value 
$\sigma^{\rm obs}_{\rm had}=27.680\pm 0.272$ nb from 
Ref.~\cite{PhysLettsB641_145_2006}.
Similarly, we determine 
the cross sections $\sigma^{\rm obs}_{\rm had}(s)$ at the other energies.
The systematic uncertainty on the cross section arises from
the track and event-selection criteria, which leads to the assignment of  
a total systematic uncertainty of $2.89\%$
for all cross sections~\cite{PhysRevLett132_191902_2024}.

We perform a least-$\chi^2$ fit to the 
observed
cross sections $\sigma^{\rm obs}_{\rm had}(s)$, which is 
modeled by
\begin{eqnarray}
\sigma^{\rm obs}_{\rm had}(s)=
f_{\rm c}\sigma^{\rm born}_{\mu^+\mu^-}(s) 
   + \int_{0}^{1- \frac{ 4{m^2_{\pi} } } {s} } dx{\sigma}^{\rm dress}_{J/\psi}(s'){\mathcal F(x,s)}~~~~~~ \nonumber  \\
+\int_{0}^{\infty}dw\mathcal G(s,w)\int_{0}^{ 1- \frac{ 4{m^2_{\pi} } } {s} } 
dx~{\sigma}^{\rm dress}_{ {\mathcal V_{>3680}} }(s'){\mathcal F(x,s)}.~~~~~~
\label{equation:Eq_observedCS}
\end{eqnarray}
In Eq.~(\ref{equation:Eq_observedCS}), 
the first term represents the observed cross section for CPLMR,
where 
$\sigma^{\rm born}_{\mu^+\mu^- }(s)$
is the born cross section for
the $e^+e^-\rightarrow \mu^+\mu^-$ continuum process and $f_{\rm c}$ 
is a free parameter. The second term represents the observed cross section
of $J/\psi\rightarrow$~hadrons including the ISR correction,
where ${\sigma}^{\rm dress}_{J/\psi}(s')$
is the dressed cross section 
(i.e. including vacuum polarization effects) 
for ${J/\psi}\rightarrow {\rm hadrons}$,
$s^{\prime}=s(1-x)$,
with $x$ a parameter relating to the total energy 
of the emitted ISR photons~\cite{Structure_Function},
$\mathcal F(x,s)$ is the sampling function~\cite{Structure_Function},
and $m_{\pi}$ is the mass of the pion.
The third term describes the observed cross sections for 
the hadronic decays of the states or structures with masses above 3.680 GeV,
${\sigma}^{\rm dress}_{\mathcal V_{>3680}}(s')={\sigma}^{\rm dress}_{\psi(3686)}(s')+{\sigma}^{\rm dress}_{X_{\rm aboveOC}}(s')$,
in which ${\sigma}^{\rm dress}_{\psi(3686)}(s')$
is the dressed cross section 
for ${\psi(3686)}\rightarrow$~hadrons,
${\sigma}^{\rm dress}_{X_{\rm aboveOC} }(s')$
is the total dressed cross section for  
$X_{\rm aboveOC}\rightarrow$~hadrons,  
including the continuum $D\bar{D}$ production, 
$\mathcal G(s,w)$ is a Gaussian function describing 
the beam-energy spread (1.355 MeV), and $w$ integrates over the energy.
The dressed cross sections for
$J/\psi$ and $\psi(3686)$ decaying into hadrons are determined by
$\sigma^{\rm dress}_{J/\psi}(s')=|{\cal A}_{J/\psi}(s')|^2$
and $\sigma^{\rm dress}_{\psi(3686)}(s')=|{\cal A}_{\psi(3686)}(s')|^2$,
where ${\cal A}_{J/\psi}(s')$ and 
${\cal A}_{\psi(3686)}(s')$
are parametrized by 
\begin{equation}
{\mathcal A_{\mathcal S}(s')} =
 \sqrt{12\pi \Gamma^{ee}_{\mathcal S} {\mathcal B} {\Gamma^{\rm tot}_{\mathcal S}} }
      /[{(s'-M_{\mathcal S}^2) + i \Gamma^{\rm tot}_{\mathcal S}M_{\mathcal S}}],
\label{equation:Eq_BW}
\end{equation}
where $\mathcal S$ indicates $J/\psi$ and $\psi(3686)$,
$M_{\mathcal S}$,  $\Gamma^{ee}_{\mathcal S}$,
and $\Gamma^{\rm tot}_{\mathcal S}$ 
are its mass, electronic width, and total width,
respectively, and $\mathcal B$ is the branching fraction for 
$\mathcal{S}\rightarrow$~hadrons.
The dressed cross section for 
$X_{\rm aboveOC}\rightarrow$~hadrons is modeled by
\begin{eqnarray}
\sigma^{\rm dress}_{{X}_{\rm aboveOC} }(s') =
 |{\cal A}_{{\rm cnt} D\bar{D}}(s') + {\cal A}_{\mathcal G(3900)}(s')e^{i\phi_0}\nonumber  \\
+ {\cal A}_{\mathcal R(3760)}(s')e^{i\phi_1}
+  {\cal A}_{\mathcal R(3780)}(s')e^{i\phi_2} \nonumber  \\
+  {\cal A}_{\mathcal R(3810)}(s')e^{i\phi_3}|^{2}.~~~~~
\label{equation:Eq_dressedCS}
\end{eqnarray}
In Eq.~(\ref{equation:Eq_dressedCS}),
the first term is the amplitude of the 
continuum $D\bar{D}$ production, 
\begin{equation}
{\cal A}_{{\rm cnt} {D\bar{D}}}(s') = \left[f_{D\bar{D}}
      \left(\beta_{0}^{3} + \beta_{+}^{3} \right)
                       \sigma^{\rm born}_{\mu^{+}\mu^{-}}(s')
                       \right]^{\frac{1}{2}},
\label{equation:Eq_cntDDb}
\end{equation}
where $\beta_0$ and $\beta_{+}$ are the 
velocities~\cite{PhysRevLetts97_121801_2006}
of the $D^0$ and $D^+$ mesons, respectively,
and $f_{D\bar{D}}$ is a free parameter;
The second term is the decay amplitude for the structure 
${\mathcal G(3900)}$~\cite{Eithtin_chmonuim_prd1980,BaBar_PRD76_111105R_Year2007}, 
for which we use
${\cal A}_{\mathcal G(3900)}(s') = [{\mathcal C}e^{{{-(\sqrt{s'}-M_{\mathcal G})^2}}/{(2\sigma^{2}_{\mathcal G})}}]^{ {\frac{1}{2}} }$
as the amplitude, where $\mathcal C$ is a free parameter, 
$M_{\mathcal G}$ and $\sigma_{\mathcal G}$ are the peak position
and standard deviation of the $\mathcal G(3900)$ structure, respectively.
The combination of the first and second terms best describe 
the shape of the nonresonant $D\bar{D}$ cross sections.
This shape is determined separately using data 
and fixed in fitting the 
measured 
$\sigma^{\rm obs}_{\rm had}(s)$ and 
their corresponding dressed cross sections below. 
We refer to this procedure as {\em cross-section spectrum separation}; 
The remaining terms: terms 
${\cal A}_{\mathcal R(3760)}(s')$, 
${\cal A}_{\mathcal R(3780)}(s')$, and 
${\cal A}_{\mathcal R(3810)}(s')$ are the amplitudes for 
$\mathcal R(3760)$, $\mathcal R(3780)$, and 
${\mathcal R(3810)}$ decaying into hadrons, respectively, 
and are described by Eq.~(\ref{equation:Eq_BW}), while
$\phi_0$, $\phi_1$, $\phi_2$ and $\phi_3$ 
in Eq.~(\ref{equation:Eq_dressedCS}) are phases.
For the $\mathcal R(3780)$, the energy dependent 
total width~\cite{PhysRevLetts97_121801_2006} is used in its amplitude.

The sum of the dressed-resonance cross sections 
and the observed-continuum-$J/\psi$ (drocj) cross section is given by
\begin{eqnarray}
{\sigma^{\rm drocj}_{\rm had}}(s) =
f_{\rm c}\sigma^{\rm born}_{\mu^+\mu^- }(s)
+\sigma^{\rm obs}_{J/\psi}(s) + {\sigma}^{\rm dress}_{{\mathcal V}_{>3680}}(s),
 \label{equation:newDressedCS_eeToHadrons}
\end{eqnarray}
where $\sigma^{\rm obs}_{J/\psi}(s)=\int_{0}^{ 1- { 4{m^2_{\pi} } }/{s}  } dx{\sigma}^{\rm dress}_{J/\psi}(s'){\mathcal F(x,s)}$.
For the study of the decays of $X_{\rm aboveOC}$ states
it is not necessary to perform radiative corrections 
for the first and second terms in
Eq.~(\ref{equation:newDressedCS_eeToHadrons}).

In the fit, the BW parameters of the $J/\psi$ and $\psi(3686)$ states
are fixed to the central values
given by the 2022 Particle Data Group~\cite{PDG2022}.
$M_{\mathcal G}$, $\sigma_{\mathcal G}$, and ${\mathcal C}$ 
are fixed to  $(3873.6\pm 4.7 \pm 0.1)$ MeV/$c^2$,
$(53.3\pm 6.3\pm 0.1)$~MeV, and 
$(1.07\pm 0.29\pm 0.03)$ nb$^{\frac{1}{2}}$,
respectively, $f_{D\bar{D}}$ is fixed to  
$(0.212\pm 0.177 \pm 0.006)$ nb$^{\frac{1}{2}}$, 
and $\phi_{0}$ is fixed to zero degrees. 
These values are determined 
by separately fitting the 
cross sections of $e^+e^-\rightarrow D\bar{D}$ at energies 
from 3.74 to 3.99 GeV, measured using the BESIII
data~\cite{data_CollectedIn2012},
in which we do not observe the $\mathcal R(3810)$;
the first uncertainties  are from the fits 
to the cross sections, and the second are due to the uncertainties 
in both the cross section measurements and the branching fraction for 
${\mathcal R(3760)}\rightarrow$~$D\bar{D}$ and
${\mathcal R(3780)}\rightarrow$~$D\bar{D}$.
In the fit, these branching fractions are fixed 
to $(85\pm 6)\%$~\cite{PhysLttesB659_74_2008},
determined by fitting cross sections for
$e^+e^-\rightarrow${non-$D\bar{D}$ over 150 energy points from 3.65 to 3.87 GeV.
We assume that the branching fractions for ${\mathcal R(3760)}\rightarrow${~hadrons},
${\mathcal R(3780)}\rightarrow${~hadrons},
and ${\mathcal R(3810)}\rightarrow${~hadrons} are all $100\%$ in the fit.
The other parameters are left free in the fit.
Using Eq.~(\ref{equation:Eq_observedCS}) 
we fit the
measured observed cross sections
$\sigma^{\rm obs}_{\rm had}(s)$ to determine the values of
the free parameters.

Inserting these parameter values into Eq.~(\ref{equation:Eq_observedCS}) and
Eq.~(\ref{equation:newDressedCS_eeToHadrons}) yields 
the expected observed cross sections 
$\sigma^{\rm obs}_{\rm had}(s)$
and expected drocj cross sections 
$\sigma^{\rm drocj}_{\rm had}(s)$,
which give the ISR correction factors
$f_{\rm ISR}(s)=\sigma^{\rm obs}_{\rm had}(s)/\sigma^{\rm drocj}_{\rm had}(s)$.
Dividing 
the measured oberved cross sections 
$\sigma^{\rm obs}_{\rm had}(s)$
by $f_{\rm ISR}(s)$ yields 
the measured
drocj cross sections 
$\sigma^{\rm drocj}_{\rm had}(s)$~\cite{Summary_of_DROCJxs_vs_Ecm}. 
The circles with error bars in
Fig.~\ref{fig:fit_DROCJ_CS_R3760Psi3770S3810} show the $\sigma^{\rm drocj}_{\rm had}(s)$ results.

Using Eq.~(\ref{equation:newDressedCS_eeToHadrons}) we fit 
the $\sigma^{\rm drocj}_{\rm had}(s)$ results 
and obtain the values for these parameters.
The nominal fit returns six solutions with fit $\chi^2$ less than 26. 
For four of the six solutions, the $\mathcal R(3780)$ electronic widths 
are less than 120 eV or larger than 415 eV, which are not consistent with 
262 eV for the $1^3D_1$-wave dominant state, i.e. $\psi(3770)$
resonance~\cite{PDG2022},
hence these are not physical solutions.
Table~\ref{tab:FitResults_CntPsi3686R3760Psi3770S3810} 
presents the values of the parameters for the remaining
two solutions, where the first uncertainties are 
from the fit to $\sigma^{\rm drocj}_{\rm had}(s)$, 
and the second are systematic. 
The two solutions have $\chi^2$ values of $25.5$ 
and $25.6$ for 29 degrees of freedom (d.o.f.).
Since no $\mathcal R(3810)$ is observed in the $D\bar{D}$ final state,
the non-open charm branching fraction of $\mathcal R(3810)$
decay must be close to $100\%$. This indicates 
$\Gamma^{ee}_{\mathcal R(3810)}=11.0\pm 2.9\pm 2.4$ eV, determined from 
$\Gamma^{ee}_{\mathcal R(3810)}\mathcal B_{\mathcal R(3810)}$ 
given in~\cite{PhysRevLett132_191902_2024}.
By comparing this value to those in 
Table~\ref{tab:FitResults_CntPsi3686R3760Psi3770S3810},
we choose solution II as the baseline result.
Compared with the measured values in Ref.~\cite{bes2_prl_2structures},
the precision of the $\mathcal R(3760)$ and $\mathcal R(3780)$
resonance parameters has improved by factors of 1.5 to 3.1.
\begin{table}
\centering
\caption{Results from the fit to the measured cross sections of $e^+e^-\rightarrow$~hadrons  
showing the values of the mass $M_{i}$ [in MeV/$c^2$], total width $\Gamma^{\rm tot}_{i}$ [in MeV], 
electronic width $\Gamma^{ee}_{i}$ [in eV], relative phase $\phi_{i}$ [in degree], and $f_{\rm c}$,
where $i$ represents $\mathcal R(3760)$, 
$\mathcal R(3780)$,
and $\mathcal R(3810)$.
}
\begin{tabular}{lcr}
\hline\hline
Parameters                                            & Solution I              &        Solution II            \\
\hline
${M}_{\mathcal R(3760)}$
                                                     & $3752.6 \pm  4.2 \pm 2.8$    & $3751.9 \pm 3.8 \pm 2.8$     \\
${\rm \Gamma^{\rm tot}_{\mathcal R(3760)} }$
                                                     & $  31.7 \pm 5.7 \pm   8.4$   &   $ 32.8 \pm 5.8 \pm 8.7$    \\
$\Gamma^{ee}_{\mathcal R(3760)}$
                                                     & $ 206 \pm 83 \pm 96$         &   $  184 \pm 75 \pm  86$     \\
$\phi_{1}$
                                                     & $  -70 \pm   23 \pm  39$     &   $   -49 \pm 29 \pm   27$   \\
${M}_{\mathcal R(3780)}$
                                                     & $3778.6 \pm  0.5 \pm 0.3$    &  $3778.7 \pm 0.5 \pm 0.3$    \\
${\rm \Gamma^{\rm tot}_{\mathcal R(3780)}}$
                                                     & $  20.3 \pm 0.8 \pm 1.7$     &   $  20.3 \pm 0.8 \pm 1.7$   \\
$\Gamma^{ee}_{\mathcal R(3780)}$
                                                     & $243 \pm 61 \pm 76$          &   $ 265 \pm 69 \pm    83$    \\
$\phi_{2}$
                                                     & $  131 \pm   16 \pm    15$   &   $  151 \pm 23 \pm   17$    \\
${M}_{\mathcal R(3810)}$
                                                     & $3804.5 \pm  0.9 \pm  0.9$   &    $3804.5 \pm 0.9 \pm  0.9$ \\
${\rm \Gamma^{\rm tot}_{\mathcal R(3810)} }$
                                                     & $  5.6 \pm 3.6 \pm  3.4$     &   $ 5.4 \pm 3.5 \pm 3.2$    \\
$\Gamma^{ee}_{\mathcal R(3810)}$
                                                     & $  2.3 \pm 0.8 \pm 1.4$      &   $ 19.4 \pm 7.4 \pm 12.1$    \\
$\phi_{3}$
                                                     & $ 81 \pm   22 \pm  18$       &   $ -11 \pm 20 \pm 2$        \\
$f_{\rm c}$                                          & $ 2.743\pm0.019 \pm 0.081$   & $2.741\pm 0.019\pm 0.081$    \\
\hline\hline
\end{tabular}
\label{tab:FitResults_CntPsi3686R3760Psi3770S3810}
\end{table}
\begin{figure}
\centering
\includegraphics[width=0.470\textwidth]{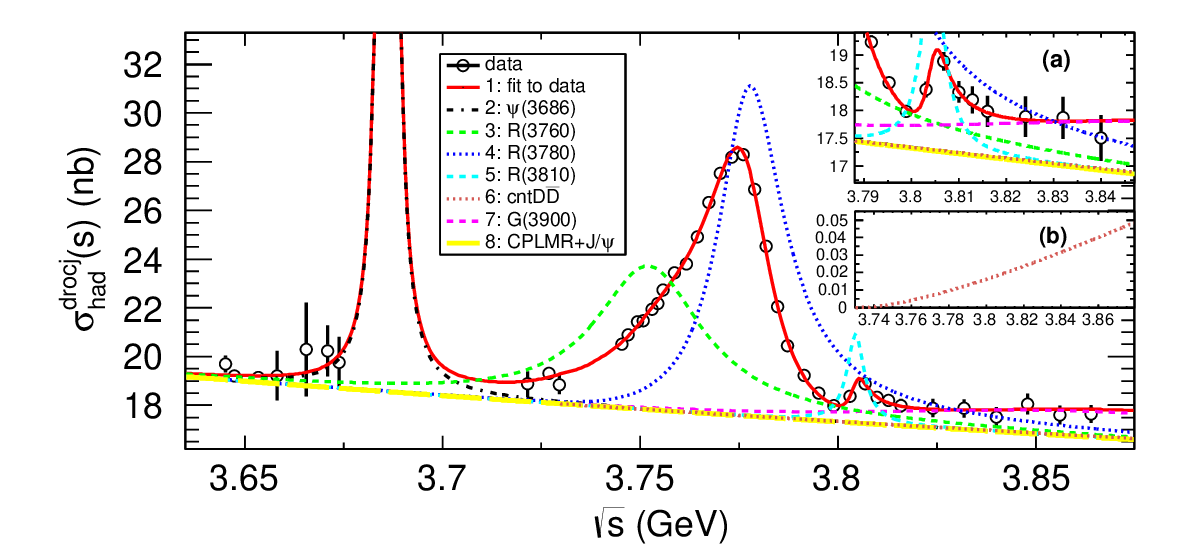}
\caption{
Measured dressed-resonance cross sections and the observed-continuum-$J/\psi$
cross sections for $e^+e^- \rightarrow {\rm hadrons}$ 
with the fit superimposed (see text for details). 
}
\label{fig:fit_DROCJ_CS_R3760Psi3770S3810}
\end{figure}

The systematic uncertainties on the values of the parameters given in
Table~\ref{tab:FitResults_CntPsi3686R3760Psi3770S3810}
originate from three sources: (i) the systematic uncertainties on the
observed cross sections, 
(ii) the uncertainties of the fixed parameters, and 
(iii) the uncertainties on the c.m.\ energies.
The estimations of these uncertainties are similar to those described in
Ref.~\cite{PhysRevLett132_191902_2024}.
Adding these uncertainties in quadrature yields the total systematic uncertainty for each
parameter value given in Table~\ref{tab:FitResults_CntPsi3686R3760Psi3770S3810}.

Figure~\ref{fig:fit_DROCJ_CS_R3760Psi3770S3810} also shows 
the fit to $\sigma^{\rm drocj}_{\rm had}(s)$, 
where the cross section described by solution II is superimposed on the results.
Also shown are the fit results including one contribution to
$\sigma^{\rm dress}_{{\mathcal R}_{{\mathcal S}>3680}}(s)$
included at a time,
which is added to the total observed 
hadronic cross sections of CPLMR and $J/\psi$.
The inset (a) shows the cross-section data in the energy range from 3.788 to
3.847 GeV, while the dashedline in (b) only shows the
cross sections of the continuum $D\bar{D}$ contribution.
The measured mass of $\mathcal R(3780)$ is consistent within $1.5\sigma$, 
and the measured mass of $\mathcal R(3760)$ and widths of $\mathcal R(3760)$ and $\mathcal R(3780)$
are consistent within $0.8\sigma$ with those~\cite{bes2_prl_2structures,sysErrOfParmts}
measured by BES-II. The values for the masses and total widths 
of the ${\mathcal R(3760)}$ and ${\mathcal R(3810)}$ measured in this work 
supersede those reported in Ref.~\cite{PhysRevLett132_191902_2024}.

To evaluate the significance of the $\mathcal R(3810)$ signal, we refit 
$\sigma^{\rm drocj}_{\rm had}(s)$ with the $\mathcal R(3760)$ 
mass and total width fixed to 
$(3739.9\pm 4.2\pm 2.6)$ MeV/$c^2$~\cite{PhysRevLett132_191902_2024}
and $(23.9\pm 8.2\pm 4.8)$ MeV~\cite{PhysRevLett132_191902_2024}, respectively.
The original fit (fit-A) including the $\mathcal R(3810)$ returns 
$\chi^2=49.7$ for 31 d.o.f.,
while the fit (fit-B) excluding the $\mathcal R(3810)$ returns 
$\chi^2=116.7$ for 35 d.o.f. 
The significance for the $\mathcal R(3810)$, including systematic uncertainty, 
is determined to be $7.4\sigma$ from the change in $\chi^2$ and number of degrees 
of freedom between the two fits.   
In an alternative approach, we refit $\sigma^{\rm drocj}_{\rm had}(s)$
with the $\mathcal R(3810)$ mass and total width fixed 
to $(3805.7\pm 1.1\pm 2.7)$ MeV/$c^2$~\cite{PhysRevLett132_191902_2024}
and $(11.6\pm 2.9\pm 1.9)$ MeV~\cite{PhysRevLett132_191902_2024}, respectively. 
The fit (fit-C) including the $\mathcal R(3760)$ returns $\chi^2=31.8$ 
for 31 d.o.f., while the fit (fit-D) excluding the $\mathcal R(3760)$
returns $\chi^2=289.8$ for 35 d.o.f. 
Comparing fit-C and fit-D, we calculate a significance for the $\mathcal R(3760)$ signal of 
$15.6\sigma$, where systematic uncertainties have been taken into account.

To investigate whether the parameterizations for the continuum $D\bar{D}$ 
amplitudes could be responsible for the cross-section enhancement
around 3.76~GeV shown in Fig.~\ref{fig:fit_DROCJ_CS_R3760Psi3770S3810}, 
we remove both the $\mathcal R(3760)$ and $\mathcal R(3810)$ amplitudes and either replace
$\sqrt{ f_{D\bar{D}} }$ in Eq.~(\ref{equation:Eq_cntDDb})
with an exponential factor
$f_{\rm NR}e^{-(p_{D^0}+p_{D^+})^2/\lambda^2}$~\cite{ExpModel_DDbCS} in the
fit (fit-E), or use a threshold function 
$(\sqrt{s} - 2m_{D^0})^{d} {e}^{-h\sqrt{s}-ts}$~\cite{BaBar_PRD76_111105R_Year2007} 
to replace ${\mathcal A}_{{\rm cnt}{D\bar{D}}}(s')$ in Eq.~(\ref{equation:Eq_dressedCS})
in the fit (fit-F), where $p_{D^0}$ ($p_{D^{+}}$) and
$m_{D^0}$ ($m_{D^+}$) are the momentum 
and mass of the $D^0$ ($D^+$) meson, respectively. Here 
$f_{\rm NR}$, $\lambda$, $d$, $h$, and $t$ 
are parameters describing the 
shape of the continuum $D\bar{D}$ contribution.
In the fits, the values for these parameters 
are fixed to those determined by analyzing the cross sections 
for $e^+e^-\!\rightarrow$~$D\bar{D}$ using the function  
given in Eq.~(\ref{equation:Eq_dressedCS}).
Fit-E and fit-F return $\chi^2=119.1$ (probability value ${\rm PV}\le 2\times 10^{-10}$) 
and $\chi^2=225.8$ (${\rm PV}\le 1\times 10^{-28}$)
for 37 d.o.f., respectively, indicating that fit-E and fit-F are strongly 
incompatible with the data.

The vector dominance model (VDM) assumes that the $\psi(3686)$ can decay into $D\bar{D}$
pairs at energies above the $D\bar{D}$ production threshold~\cite{PRD81_034011_ZQiang}.
To check the validity of the VDM approach, we remove the amplitude 
${\cal A}_{\mathcal R(3810)}(s')$, and
replace the amplitude ${\cal A}_{\mathcal R(3760)}(s')$ in Eq.~(\ref{equation:Eq_dressedCS}) with  
${\mathcal A_{\psi'}}(s') = \sqrt{12\pi \Gamma^{ee}_{\psi'}\Gamma^{D\bar D}_{\psi'}}/[(s'-M_{\psi'}^2)+i\Gamma^{\rm
{tot~aboveD\bar{D}}}_{\psi'}M_{\psi'}(s')]$~\cite{KEDR}
in the fit (fit-G), where $\psi'$ is the $\psi(3686)$,
while $\Gamma^{\rm {tot~above{D\bar D} }}_{\psi'}$
and $\Gamma^{D\bar{D}}_{\psi'}$ are the total and $D\bar{D}$ 
widths of this resonance, respectively. $\Gamma^{\rm {tot~above{D\bar D}}}_{\psi'}$ is 
chosen to be energy dependent and defined as that for 
$\psi(3770)$~\cite{PhysRevLetts97_121801_2006,PhysLettsB652_238_2007}
with branching fractions 
$\mathcal B_{00}\![\psi(3686)\!\rightarrow$~$D^0\bar{D^0}]$
and $\mathcal B_{+-}\![\psi(3686)\!\rightarrow$~$D^+D^-]$
fixed to $56\%$ and $44\%$, respectively. In fit-G, the values 
for $\Gamma^{\rm {tot~above{D\bar D} }}_{\psi'}$, 
$\Gamma^{D\bar{D}}_{\psi(3686)}$ and $\phi_1$ 
are fixed to those determined by analyzing the cross sections 
for $e^+e^-\!\rightarrow$~$D\bar{D}$ using
the function given in Eq.~(\ref{equation:Eq_dressedCS}).
Fit-G returns $\chi^2=139.4$ (${\rm PV}\le 9\times 10^{-14}$)
for 37 d.o.f.. 
This poor goodness of fits indicates that the VDM approach 
alone cannot explain the $\mathcal R(3760)$ and $\mathcal R(3810)$ 
structures~\cite{RulOutSomeExplanations}.

The {\em cross-section spectrum separation}
technique performed in fitting these $\sigma^{\rm drocj}_{\rm had}(s)$ 
accurately separates the final states of $\mathcal R(3760)$, $\mathcal R(3760)$, 
and $\mathcal R(3810)$ decays to hadrons from the final states 
of nonresonant decay to $D\bar{D}$, thereby achieving 
the most precise measurement of the parameters of the three resonances. 
The results returned from fitting $\sigma^{\rm drocj}_{\rm had}(s)$ 
make clear that the $\psi(3770)$ is a complex system composed of the $\mathcal R(3760)$, 
$\mathcal R(3780)$, and $\mathcal R(3810)$ resonances. 
In other words, the $\psi(3770)$~\cite{LGW,DLCO} resonance 
is a single BW artifact that is compatible with less precise measurements of the  cross section
of $e^+e^-\rightarrow$~ hadrons at energies
from 3.73 to 3.87 GeV, where the three BW 
structures [labeled as $\mathcal R(3760)$, $\mathcal R(3780)$, 
and $\mathcal R(3810)$ in the Letter]
cannot be distinguished. Therefore, the $\psi(3770)$ cannot itself be 
the $1^3D_1$-wave dominate state of charmonium.

The potential model~\cite{Eithtin_chmonuim_prd1978} predicts that 
only the charmonium ${1}^3D_1$ state
can be produced in $e^+e^-$ collisions  
at c.m.\ energies from 3.732 to 3.871 GeV.
A model called "a new spectroscopy"
~\cite{A_De_Rujule_PRL38_317_Year1977}
proposes that a $p$-wave resonance of a four-quark ($c\bar{c}q\bar{q}$) state
can be produced in $e^+e^-$ collisions 
at c.m.\ energies above 3.732 GeV.
This four-quark state can be either a simple four-quark bound state or
an open charm molecular state. It is natural to interpret 
the $\mathcal R(3760)$ as an open charm pair molecular 
state~\cite{Dubynskiy_Voloshin_PRF78_116014_2008}, since its mass of 
$(3751.9\pm 3.8 \pm 2.8)$ MeV/$c^2$ is $(12.6\pm{4.7})$ MeV/$c^2$
higher than the $D^+{D^-}$ production threshold $(3739.3\pm 0.1)$ MeV/$c^2$
and its total width is $(32.8\pm 5.8 \pm 8.7)$ MeV.
As the $\mathcal R(3760)$ is also observed in nonopen charm hadron final
states with a resonance peak at 
$(3739.7\pm 4.7)$ MeV/$c^2$~\cite{PhysRevLett132_191902_2024},
the $\mathcal R(3760)$ could also contain a component of the simple four-quark state. 
This feature is consistent with the four-quark state production 
and decay~\cite{A_De_Rujule_PRL38_317_Year1977}.
The $\mathcal R(3810)$ can be interpreted as a hadro-charmonium 
state~\cite{Voloshin_PLB66_344_Year2008}, 
since no $\mathcal R(3810)$ is observed in the cross sections of
$e^+e^-\rightarrow D\bar{D}$, and its mass, 
which is $(3804.5\pm 0.9 \pm 0.9)$ MeV/$c^2$,
is very close to the $h_c\pi^+\pi^-$ threshold $(3804.5\pm 0.1)$ MeV/$c^2$.
The $\mathcal R(3780)$ total width, which is $(20.3\!\pm0.8\!\pm\!1.7)$ MeV,  
is consistent with the range of possible values from 10 to 26~MeV predicted by the potential 
models~\cite{Heikkila_PRD29_110_1984,SOno_PRD23_1118_1981,Eichten_PRD69_094019_Y2004}
for the $1^3D_1$ state.
The potential model~\cite{Eichten_PRD69_094019_Y2004} also predicts that
the total width of $1^3D_1$ state is less (larger) than 7 (47) MeV 
for a mass of 3752 (3805) MeV,
indicating that neither the $\mathcal R(3760)$ 
nor the $\mathcal R(3810)$ is the $1^3D_1$ state. 
These experiment results and theoretical predictions lead 
us to interpret the $\mathcal R(3780)$ as a resonance dominated by the
$1^3D_1$-wave charmonium state.

In summary, measurements of the total cross sections 
for $e^+e^- \rightarrow$~hadrons~\cite{Ruds_R} at c.m.\ energies 
from 3.645 to 3.871 GeV have been performed with higher precision than in previous analyses. 
We have performed the most precise measurements of the parameters of 
the $\mathcal R(3760)$, $\mathcal R(3780)$, and $\mathcal R(3810)$ resonances.
The $\mathcal R(3810)$ can be interpreted as a hadro-charmonium state. 
The $\mathcal R(3760)$ can be interpreted as an 
open-charm molecular state, but may contain a component from a four-quark state contribution.
We have established that the $\psi(3770)$ is not a single resonance, but a complex system composed
of the $\mathcal R(3760)$, $\mathcal R(3780)$, and $\mathcal R(3810)$ resonances.
Among these, we interpret the $\mathcal R(3780)$ is a
resonance dominated by the $1^3D_1$ charmonium state.
These results provide important experimental input 
to help in the understanding of the natures of $c\bar{c}$ and non-$c\bar{c}$ states.

{\em Acknowledgments---}The BESIII Collaboration 
thanks the staff of BEPCII and the IHEP computing center for their strong support. 
This work is supported in part by 
National Key R$\&$D Program of China under Contracts No. 
2009CB825204, No. 2020YFA0406300, No. 2020YFA0406400; 
National Natural Science Foundation of China (NSFC) under Contracts No. 
10935007, No. 11635010, No. 11735014, No. 11835012, No. 11935015, No. 11935016, 
No. 11935018, No. 11961141012, No. 12022510, No. 12025502, No. 12035009, No. 12035013, 
No. 12061131003, No. 12192260, No. 12192261, No. 12192262, No. 12192263, No. 12192264, 
Np. 12192265, No. 12221005, No. 12225509, No. 12235017; 
the Chinese Academy of Sciences (CAS) Large-Scale Scientific Facility Program; 
the CAS Center for Excellence in Particle Physics (CCEPP); 
CAS Key Research Program of Frontier Sciences under Contracts No. 
QYZDJ-SSW-SLH003, No. QYZDJ-SSW-SLH040; 100 Talents Program of CAS; 
the CAS Research Program under Code No. Y41G1010Y1;
the CAS Other Research Program under Code No. Y129360;
The Institute of Nuclear and Particle Physics (INPAC) and Shanghai Key Laboratory for Particle Physics and Cosmology; 
ERC under Contract No. 758462; European Union's Horizon 2020 research 
and innovation programme under Marie Sklodowska-Curie grant agreement under Contract No. 894790; German Research Foundation DFG 
under Contracts No. 443159800, No. 455635585, Collaborative Research Center CRC 1044, FOR5327, GRK 2149; 
Istituto Nazionale di Fisica Nucleare, Italy; 
Ministry of Development of Turkey under Contract No. DPT2006K-120470; 
National Research Foundation of Korea under Contract No. NRF-2022R1A2C1092335; 
National Science and Technology fund of Mongolia; 
National Science Research and Innovation Fund (NSRF) via the Program Management Unit for Human Resources  
$\&$
Institutional Development, 
Research and Innovation of Thailand under Contract No. B16F640076; Polish National Science Centre under Contract No. 2019/35/O/ST2/02907; 
The Swedish Research Council; U. S. Department of Energy under Contract No. DE-FG02-05ER41374


\begin{thebibliography}{99}
\bibitem{sysErrOfParmts}
The systematic uncertainties of the fitted parameters mainly arise from the
uncertainties of the fixed parameter values, which lead to the systematic uncertainties 
of the fitted parameters being more than 5 times greater than those
measured by the BES-II~\cite{bes2_prl_2structures}.
In the BES-II measurements, 
no parameter value is fixed when fitting the cross sections 
for $e^+e^-\rightarrow${hadrons}.
\end{thebibliography}
\end{document}